\documentclass[twocolumn,amssymb,superscriptaddress,aps,prd,floatfix]{revtex4}
\usepackage{amsmath}
\usepackage[dvips]{graphicx}
\usepackage[dvips]{color}
\usepackage[usenames,dvipsnames]{xcolor}
\newlength{\colw}
\newcommand{\One}{1\kern-4.5pt1}

\definecolor{DarkGreen}{rgb}{0,0.5,0}

\begin{document}

\title{Further Evidence For Zero Crossing On The Three Gluon Vertex}

\author{Anthony G. Duarte}
\affiliation{CFisUC, Department of Physics, University of Coimbra, P-3004 516 Coimbra, Portugal} 

\author{Orlando Oliveira}
\affiliation{CFisUC, Department of Physics, University of Coimbra, P-3004 516 Coimbra, Portugal} 

\author{Paulo J.\ Silva}
\affiliation{CFisUC, Department of Physics, University of Coimbra, P-3004 516 Coimbra, Portugal}

\begin{abstract}
The three gluon one particle irreducible function is investigated using lattice QCD simulations
over a large region of momentum in the Landau gauge for four dimensional pure Yang-Mills equations and the SU(3)
gauge group. The results favor a zero crossing of the gluon form factor for momenta in the range $220 - 260$ MeV. This zero crossing 
is required to happen in order to have a properly defined set of Dyson-Schwinger equations. 
It is also shown that in the high momentum region the lattice results are compatible
with the predictions of renormalisation group improved perturbation theory.
\end{abstract}

\pacs{11.15.Ha,12.38.Aw,21.65.Qr}
         
\maketitle

The non-perturbative computation of the QCD Green's functions is an important step towards the understanding 
of the dynamics of strong interactions. The calculation of the Green's functions over the full momentum range 
is a non-trivial task, independently of taking into account or not the contribution of quarks.  
The lattice approach to QCD allows for first principles determination of the $n$-point complete Green functions as for example
\begin{equation}
   G^{(n)} (x_1, \dots , x_n )  = \langle A(x_1) \cdots A(x_n) \rangle \ ,
\end{equation}   
where $A$ stands for the gluon field and where we have omitted the color and Lorentz indices. In momentum space the Green's functions reads
\begin{equation}
   G^{(n)} (p_1, \dots , x_n )  = \langle A(p_1) \cdots A(p_n) \rangle \ .
   \label{Eq:npointGreen}
\end{equation}   
The complete Green's functions $G^{(n)}$ can be decomposed in terms of one particle irreducible (1PI) functions $\Gamma^{(n)}$ and, therefore,
from $G^{(n)}$ one can have access to the various form factors or combination of form factors that define $\Gamma^{(n)}$.

For a quantum field theory the knowledge of the 1PI enables the computation of any Green's function, of the matrix S and
the determination of the cross sections predicted by the theory. Furthermore, the 1PI functions summarise, like the Green's functions, the
dynamics of the theory and for QCD they encode information on quark and gluon confinement. 
For gauge theories the Green's functions and the 1PI are gauge dependent objects and, therefore, one has to choose 
\textit{a priori} a given gauge.

In this paper, we will report on lattice QCD results for the SU(3) pure Yang-Mills theory in the minimal Landau gauge. The focus
will be on the momentum space three point Green's function $G^{(3)} \, ^{a_1a_2a_3}_{{\mu}_1{\mu}_2{\mu}_3} (p_1, p_2,p_3)$ 
and the corresponding 1PI  function $\Gamma^{(3)} \, ^{a_1a_2a_3}_{{\mu}_1{\mu}_2{\mu}_3} (p_1, p_2,p_3)$ . 

The computation of the three point Green's function is connected with multiple properties of the strong interactions. For example,
from $G^{(3)} $ one can define a static potential between color charges or measure the strong coupling 
constant~\cite{Parrinello1994,Alles1997,Boucaud2000,Soto2001,Alkofer2005,Kellermann2008,Cornwall2012,Boucaud2014,Athenodorou2016}, 
an important phenomenological input for studying strong interaction physics. 

The three point function is also a crucial ingredient in the continuum approach to strong interactions. Indeed, the reproduction of the 
lattice results for the gluon and ghost propagators using the Dyson-Schwinger equations requires a proper description of this correlation 
function~\cite{Binosi2011,Huber2013}. Furthermore, in the continuum approach to non-perturbative QCD, the three gluon vertex 
plays a fundamental role in the structure of the Dyson-Schwinger equations (DSE). Indeed, given that the ghost propagator is 
described by an essentially massless particle type of propagator, the ghost loops in the DSE give rise to a new type of divergences which, 
to be removed and to define a finite set of equations, introduce non-trivial constraints into the structure of the Green's functions with 
a higher number of external legs.

For the particular case of the pure Yang-Mills theory and for the gluon Dyson-Schwinger equation, assuming an essentially tree level like ghost propagator,
and 
assuming that the four-gluon vertex is subleading in the infrared region, the requirement of a finite DSE for the gluon requires
that certain form factors associated with the three gluon 1PI change sign for momentum in the IR region~\cite{Binosi2013,Aguilar2014} and, therefore,
they become zero at certain kinematical points.
Similar requirements on $\Gamma^{(n)}$ for $n > 3$ also appear when one considers the DSE associated to the Green functions
with a larger number of external legs~\cite{BinosiIbanez2014}. In this sense, the cancelation of the divergences associated to
the ghost loop in the DSE provides a highly non-trivial constrain on 1PI at sufficiently small momenta. 

In what concerns the three gluon vertex in pure Yang-Mills theory, the zero crossing of the form factors describing the 1PI
has been observed in lattice simulations in three space-time dimensions but not in four dimensional simulations for the gauge group 
SU(2)~\cite{Cucchieri2006,Cucchieri2008}. For SU(3) pure Yang-Mills in four dimensions,
the change of sign was reported very recently in~\cite{Athenodorou2016}.
Furthermore, the change of sign was also seen in the solutions of the Dyson-Schwinger equations 
for the three gluon vertex~\cite{Blum2014,Eichmann2014}, on the
variational solution of QCD in the Coulomb gauge~\cite{Campagnari2010} and on the description of the Yang-Mills theory via the 
Curci-Ferrari model~\cite{Pelaez2013}. Although the momentum scale where the zero crossing in the three gluon vertex sets in
is not exactly the same in all the works just mentioned, it seems to be a common feature of the modern 
description of the three gluon vertex. In three dimensional SU(2) lattice simulations~\cite{Cucchieri2008} the momentum
scale where the form factors become negative was estimated as a value in  the range $\sim 150$ -- $250$ MeV.
The study of the gluon DSE~\cite{Aguilar2014} reports a momentum scale of $\sim 130$ -- $200$ MeV where the
change of sign for the three gluon vertex occurs. The lattice simulation for SU(3) points towards a momentum scale that is clearly above 
100 MeV~\cite{Athenodorou2016}. 
As investigated in~\cite{ Williams2016}, the inclusion of the dynamical effects in the three gluon vertex confirms the pattern described above
with the zero crossing happening within the same scales of momentum as in the quenched theory.

In the current paper we report on SU(3) lattice estimations of the three gluon vertex for pure Yang-Mills theory in the minimal Landau gauge. 
Our aim is to provide further evidence that the form factors associated to this 1PI function vanish for certain kinematical configurations
and deliver information which can help to parametrise the three gluon 1PI in the continuum approach to QCD.

This work is organised as follows. In Sec. \ref{Sec:Def} we resume the lattice setup and the definitions used throughout the paper.
In Sec. \ref{Sec:Green} we describe the various type of Green functions considered here and, in particular, the one particle irreducible function
associated to the three gluon vertex. In Sec. \ref{Sec:resultados} the lattice simulation results are reported for  two different lattices.
Finally, in Sec. \ref{Sec:Final} we summarize and conclude. In App. \ref{Sec:GammaBallChiu}, for completeness,  
we provide a tensor decomposition of the three gluon vertex.

%===================================================================================
%===================================================================================
\section{Lattice Setup and Definitions \label{Sec:Def}}

The simulations reported here use the standard Wilson action, which reproduces the continuum action up to corrections of order $a^2$.
For the generation of the gauge configurations we used the Chroma library~\cite{Chroma} and a combined Monte Carlo sweep of
seven over-relaxation updates with four heat bath updates. The gauge configurations generated by importance sampling
were rotated to the Landau gauge by
maximising the function
\begin{equation}
   F_U[g] = \frac{1}{V N_d N_c} \sum_{x , \mu} \mbox{Re Tr} \left[ g(x) \, U_\mu (x) \, g^\dagger (x + a \hat{e}_\mu) \right] \ ,
\end{equation}
over each gauge orbit associated to the link $U_\mu (x)$,
where $V$ is the number of the lattice points, $N_d = 4$ is the number of space-time dimensions, $N_c = 3$ is the number of colours
and $\hat{e}_\mu$ is the unit vector along the lattice direction $\mu$. 
The gauge fixing, i.e. the maximisation of $F_U[g]$, was performed with the Fourier accelerated steepest descent
method described in~\cite{GaugeFixing}, which was implemented using Chroma and  PFFT~\cite{PFFT} libraries. 
The quality of the gauge fixing was monitored with
\begin{equation}
\theta = \frac{1}{V N_c} \sum_x \mbox{Tr} \left[ \Delta(x) \Delta^\dagger(x) \right]
\end{equation}
where
\begin{equation}
 \Delta (x) = \sum_\mu \left[  U_\mu (x - \hat{e}_\mu) -  U_\mu (x) -  ~ h.c. ~ -  ~ trace \right]  \ ,
\end{equation}  
the lattice version of the Landau gauge fixing condition $\partial A (x) = 0$. 
For each gauge configuration, the gauge fixing was stopped when $\theta \leqslant 10^{−15}$.

For the computation of the complete Green's functions we take the definitions as those in~\cite{Silva2004}.
The gluon field is given by
\begin{multline}
a \, A_\mu (x + a \hat{e}_\mu)  = \frac{ U_\mu (x) - U^\dagger (x)}{ 2 i g_0} \\
      - \frac{ \mbox{Tr} \left[ U_\mu (x) - U^\dagger (x) \right]}{6 i g_0} + \mathcal{O}(a^2) \ ,
\end{multline} 
which reads in momentum space as
\begin{equation}
 A_\mu (\hat{p}) = \sum_x e^{- i \hat{p} (x + a \hat{e}_\mu) } \, A_\mu (x + a \hat{e}_\mu) \ ,
\end{equation}
where
\begin{equation}
   \hat{p}_\mu = \frac{2 \, \pi \, n_\mu}{a \, L_\mu} \qquad\mbox{and}\qquad n_\mu = 0, \dots, L_\mu
\end{equation}
and $L_\mu$ is the number of lattice points in direction $\mu$. The results reported are given in terms of
the tree level improved momentum
\begin{equation}
   p_\mu = \frac{2}{a} \, \sin \left( \frac{a \,  \hat{p}_\mu}{2} \right) \ .
   \label{Eq:DefMom}
\end{equation}   
Furthermore, to minimise lattice artefacts associated to the break of the rotational symmetry, for momenta above 1 GeV
we only report the data surviving to the momentum cuts as defined in~\cite{Leinweber1998}.

For the evaluation of the statistical errors,  the results reported here rely on the bootstrap method using a confidence level of
67.5\%.

%==================================================================
%==================================================================
\section{Green Functions and 1PI Functions \label{Sec:Green}}

Our goal is to describe the three gluon vertex which requires the computation of the complete Green function (\ref{Eq:npointGreen}) for
two and three external legs. On the lattice, the two point function is given by
\begin{equation}
    \langle A^{a_1}_{\mu_1} (p_1) \, A^{a_2}_{\mu_2} (p_2) \rangle = V \, \delta( p_1 + p_2 ) ~ D^{a_1 a_2}_{\mu_1 \mu_2} (\hat{p}_1)
\end{equation}
where, in the Landau gauge, the gluon propagator reads
\begin{equation}
   D^{a_1 a_2}_{\mu_1 \mu_2} (\hat{p}) = \delta^{a_1 a_2} ~ P_{\mu_1 \mu_2} (p) ~ D(p^2_1)
\end{equation}
and the orthogonal projector is given by
\begin{equation}
   P_{\mu_1 \mu_2} (p) = \delta_{\mu_1 \mu_2} - \frac{p_{\mu_1} p_{\mu_2}}{p^2} \ .
\end{equation}

The three point complete Green's function has a similar definition
\begin{multline}
    \langle A^{a_1}_{\mu_1} (p_1) \, A^{a_2}_{\mu_2} (p_2) \, A^{a_3}_{\mu_3} (p_3) \rangle =  \\ V \, \delta( p_1 + p_2 + p_3) ~
   G^{a_1 a_2 a_3}_{\mu_1 \mu_2 \mu_3} (p_1, p_2, p_3) 
   \label{Eq:GLat}
\end{multline}
and it can be written in terms of the gluon propagator and of the 1PI vertex as
\begin{multline}
  G^{a_1a_2a_3}_{\mu_1\mu_2\mu_3} (p_1, p_2, p_3) =  
   D^{a_1b_1}_{\mu_1\nu_1}(p_1) ~ D^{a_2b_2}_{\mu_2\nu_2}(p_2) ~ D^{a_3b_3}_{\mu_3\nu_3}(p_3) \\
   \Gamma^{b_1b_2b_3}_{\nu_1\nu_2\nu_3} (p_1, p_2, p_3) \ ,
\end{multline}
where there is an implicit sum over the  $b_i$ and $\nu_i$ indices.
The color structure of the three gluon vertex allows one to write the 1PI as
\begin{equation}
  \Gamma^{a_1 a_2 a_3}_{\mu_1 \mu_2 \mu_3} (p_1,  p_2, p_3) = f_{a_1 a_2 a_3} \Gamma_{\mu_1 \mu_2 \mu_3} (p_1, p_2, p_3) \, ,
  \label{Eq:gammaLorentz}
\end{equation}  
where $f_{abc}$ stands for the SU(3) group structure constants. Bose symmetry requires the vertex to be symmetric under the interchange of any pair $(p_i, a_i, \mu_i)$ and  it follows that $\Gamma_{\mu_1 \mu_2 \mu_3} (p_1, p_2, p_3)$ must be antisymmetric under interchange of any pair of $(p_i, \mu_i)$. 

In order to describe the $\Gamma_{\mu_1 \mu_2 \mu_3} (p_1, p_2, p_3)$ one has to define a given Lorentz tensor basis of operators. Its
general form in the continuum has been investigated in~\cite{BallChiu1980} and decomposes the 1PI function in terms of
transverse $\Gamma^{(t)}$ and longitudinal $\Gamma^{(l)}$ components. The function $\Gamma_{\mu_1 \mu_2 \mu_3} (p_1, p_2, p_3)$ can be written
in terms of six Lorentz invariant form factors with two form factors associated to $\Gamma^{(t)}$ and the remaining associated to $\Gamma^{(l)}$.
The details of the decomposition in terms of the form factors can be found in appendix~\ref{Sec:GammaBallChiu}.
 
Our starting point to access the 1PI three gluon vertex is the following color trace
\begin{widetext}
\begin{multline}
G_{\mu_1 \mu_2 \mu_3} (p_1, p_2, p_3 ) = 
 \mbox{Tr} ~ \langle A_{\mu_1} (p_1) \, A_{\mu_2} (p_2) \, A_{\mu_3} (p_3)    \rangle = \\
 = \frac{N_c(N^2_c-1)}{4} ~ D(p^2_1) \, D(p^2_2) \, D(p^2_3) ~
 P_{\mu_1\nu_1}(p_1) \,  P_{\mu_2\nu_2}(p_2) \, P_{\mu_3\nu_3}(p_3) ~ \Gamma_{\nu_1 \nu_2 \nu_3} (p_1, p_2, p_3)
 \label{Eq:tracecompleto}
\end{multline}
\end{widetext}
where, to simplify the above expression, we have omitted the factor $V \delta(p_1 + p_2 + p_3)$ and 
where $\langle \cdots \rangle$ means average over  gauge configurations. 
We call the reader's attention, that in previous lattice calculations of the three gluon 1PI for the gauge groups SU(2) and SU(3)
the authors considered a different function, named generically $R$, which is given by a ratio of two and three point correlation functions; 
see~\cite{Cucchieri2006,Cucchieri2008,Athenodorou2016} for details.

 %================================================================
 %================================================================
 \section{Results \label{Sec:resultados}}

\begin{figure*}[t] %  figure placement: here, top, bottom, or page
   \centering
   \includegraphics[width=3.3in]{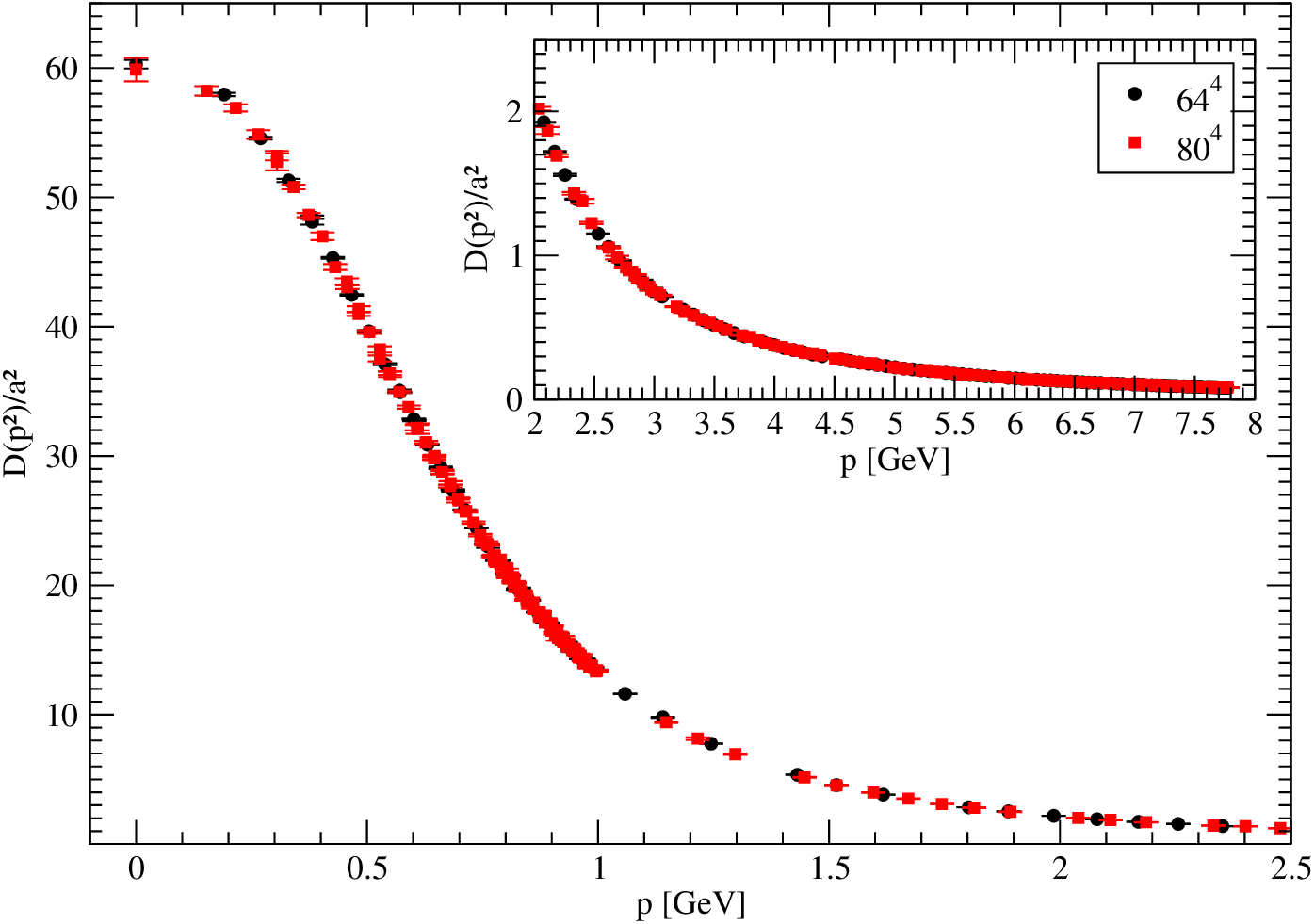}  \hfill
   \includegraphics[width=3.3in]{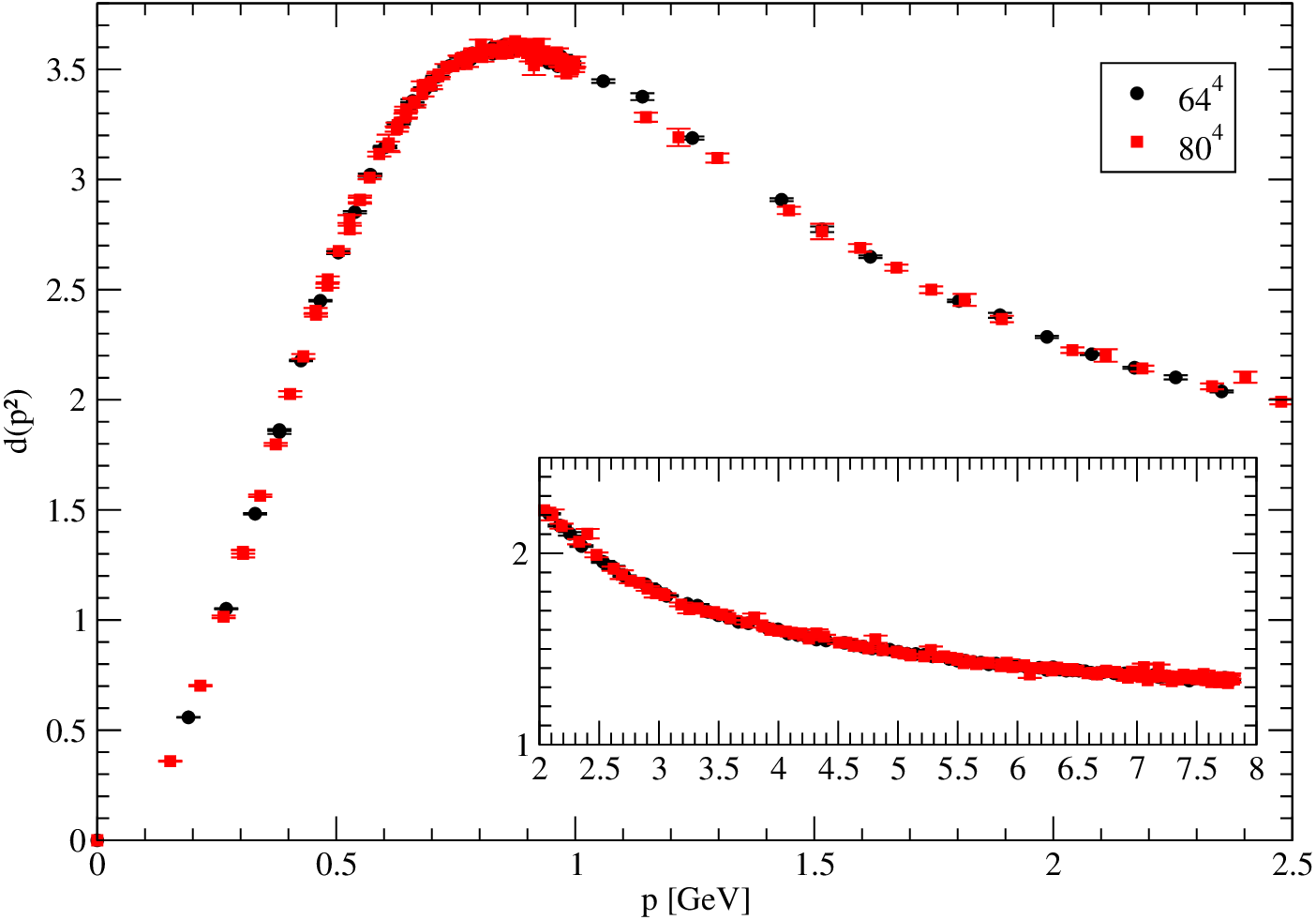} 
   \caption{Bare Landau gauge gluon propagator (left) and dressing function $d(p^2) = p^2 D(p^2)$ (right).}
   \label{fig:prop}
\end{figure*}

 \begin{figure}[t] %  figure placement: here, top, bottom, or page
   \centering
   \vspace*{0.5cm}
   \includegraphics[width=3.3in]{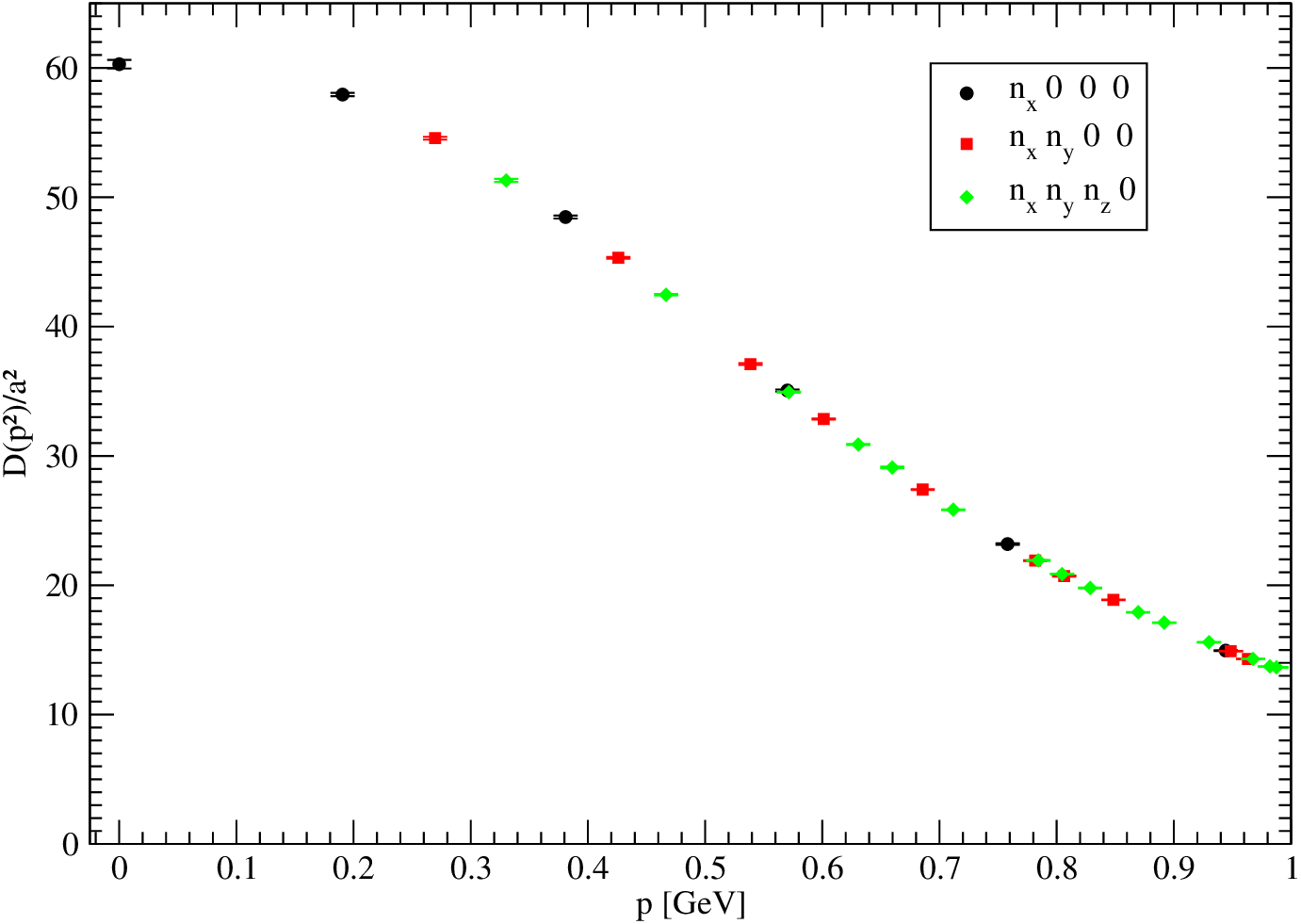} 
   \caption{Infrared bare Landau gauge gluon propagator for different types of momenta computed using the $64^4$ lattice.}
   \label{fig:propXXX}
\end{figure}

The lattice simulations can only measure the Green's functions on a limited set of kinematical configurations. Herein, we report the results
of the simulations performed using $64^4$ and $80^4$ lattices and for $\beta = 6.0$. For the conversion into physical units we take the lattice spacing
as measured from the string tension~\cite{Bali1993} and, therefore, $1/a = 1.943$ GeV or $a = 0.102$ fm. The smaller lattice has a physical volume
of $(6.5$ fm$)^4$. For the smallest lattice, the smallest momentum is $p_{min} = 191$ MeV and we generated 2000 gauge configurations. 
The largest lattice has a physical volume of $(8.2$ fm$)^4$, the smallest momentum being $p_{min} = 153$ MeV and we use 279 gauge configurations
in the computation of all correlation functions.

In Fig.~\ref{fig:prop} the bare Landau gauge gluon propagator and dressing function $d(p^2) = p^2 D(p^2)$ are reported
as a function of momenta. For $p$ above 1 GeV the data plotted includes only those momenta which satisfy the cuts as defined in~\cite{Leinweber1998}.
For momenta below 1 GeV we have included all lattice data points.
No clear finite size and volume effects are seen in the gluon propagator data. Indeed, for the smaller momenta, if we plotted 
$D(p^2)$ separating the different types of momenta, i.e. the $(n_x \, 0 \, 0 \, 0)$ momenta from $(n_x \, n_y \, 0 \, 0)$ and from $(n_x \,  n_y \, n_z \, 0)$, 
the bare lattice data seems to define a unique curve. The propagator for the different types of momenta can be seen in Fig.~\ref{fig:propXXX}
for the $64^4$ lattice. 
Although, at the level of the gluon propagator there seems to be no finite size effects, the same does not apply to the
computation of the three point function, as seen below.

Let us consider the case where one of the gluon momenta vanishes, i.e. $p_2 = 0$. This particular kinematical configuration was investigated
in the first lattice study of the three gluon vertex~\cite{Alles1997} and was used to compute the strong coupling constant.
For a gluon with $p = 0$, the momentum projector operator simplifies and is given by $P_{\mu\nu} (0) = \delta_{\mu\nu}$. A straightforward calculation
shows that the continuum complete three point Green function requires only the knowledge of two longitudinal form factors and
\begin{multline}
   G_{\mu_1\mu_2\mu_3} (p, 0, -p)  =     \left[D(p^2)\right]^2 \, D(0) \\ \frac{\Gamma (p^2)}{3} ~ ~ p_{\mu_2} ~T_{\mu_1\mu_3} (p) 
   \label{Eq:G3basico}
\end{multline}   
where $\Gamma (p^2)$ is given in terms of the invariant form factors introduced in~\cite{BallChiu1980} as
\begin{equation}
  \Gamma (p^2)   =  2 \bigg[ A(p^2,p^2;0) +  p^2 \, C(p^2,p^2;0) \bigg] \ ;
\end{equation}   
see also appendix~\ref{Sec:GammaBallChiu}. Note that there is an overall minus sign relative to the expressions reported in appendix~\ref{Sec:GammaBallChiu} which is due to the rotation to the Euclidean space. 
In the following, we will report the form factor $\Gamma (p^2)$
as measured from the combination
\begin{multline}
   G_{\mu \, \alpha \,\mu} (p, 0, -p) \, p_\alpha = 
   \, \left[D(p^2)\right]^2 \, D(0) ~~\Gamma (p^2) ~~ p^2 \ ,
   \label{Eq:Ffactorp0p}
 \end{multline}
 where we use the momentum definition as given in Eq, (\ref{Eq:DefMom}).
We recall the reader that the color factor $N(N^2-1)/4$ appearing in (\ref{Eq:tracecompleto}) was omitted in the above expression. 

For the gauge group SU(2), the authors~\cite{Cucchieri2006,Cucchieri2008} considered instead of $\Gamma (p^2)$,
the contraction of the complete Green function with the lattice tree level tensor structure $\Gamma^{(0)}$, i.e. the function
\begin{widetext}
\begin{equation}
 R(p^2) =  \frac{\Gamma^{(0)\, a_1 a_2 a_3}_{{\mu_1} \, {\mu_2} \, {\mu_3}} (p, 0, -p)  ~ \Gamma^{a_1 a_2 a_3}_{{\mu_1} \, {\mu_2} \, {\mu_3}} (p, 0, -p)}
                        {\Gamma^{(0)\, a_1 a_2 a_3}_{{\mu_1} \, {\mu_2} \, {\mu_3}} (p, 0, -p) D^{a_1 b_1}_{\mu_1 \nu_1} (p)D^{a_2 b_2}_{\mu_2 \nu_2} (0)
                                               D^{a_3 b_3}_{\mu_3 \nu_3} (p) \Gamma^{(0)\, b_1 b_2 b_3}_{{\nu_1} \, {\nu_2} \, {\nu_3}} (p, 0, -p) } \ .
\end{equation}
\end{widetext}
The lattice tree level $\Gamma^{(0)}$ can be found in e.g.~\cite{Rothe}.
A straightforward calculation using the decomposition reported in App.~\ref{Sec:GammaBallChiu}, and
assuming the continuum expressions for the various quantities, shows that $R(p^2) = \Gamma (p^2) /2$.

\begin{figure*}[t] %  figure placement: here, top, bottom, or page
   \centering
   \includegraphics[width=3.3in]{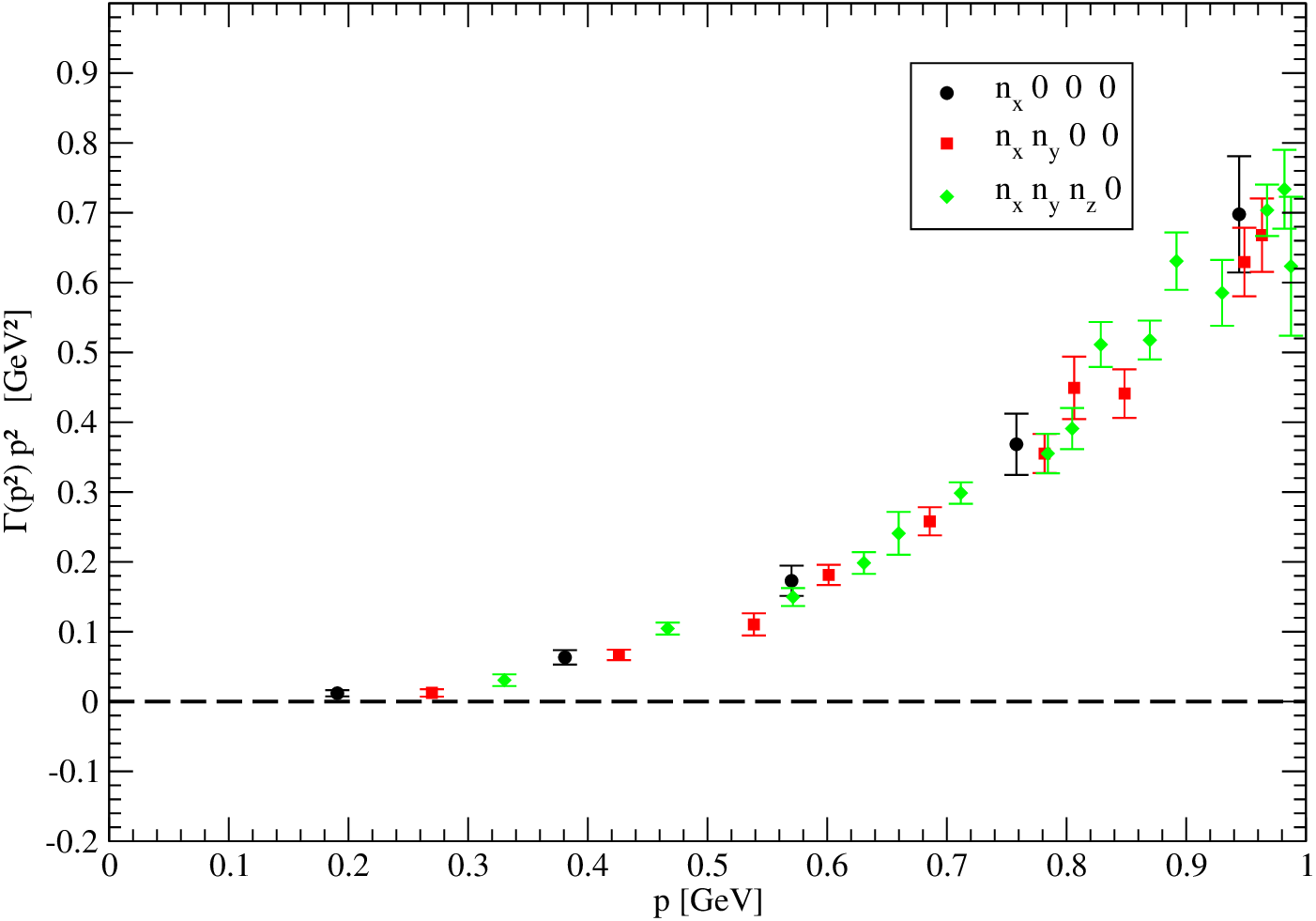} 
   \hfill
   \includegraphics[width=3.3in]{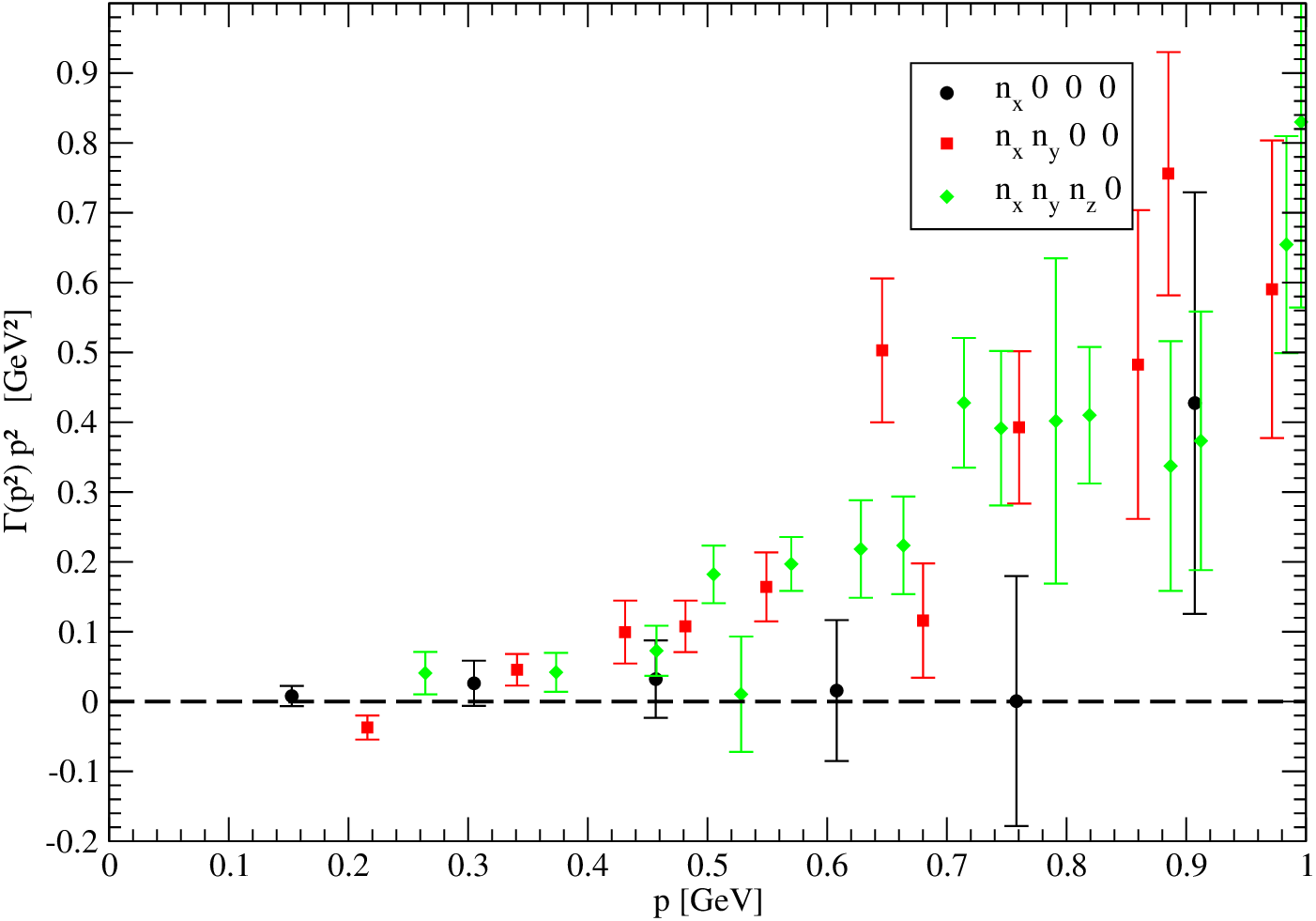} 
   \caption{Infrared $\Gamma (p^2) \, p^2$ computed using the $64^4$ (left) and $80^4$ (right) data sets for different types of momenta.}
   \label{fig:1PIXXX}
\end{figure*}

The measurement of the form factor $\Gamma (p^2)$ requires the computation of the ratio
$ G_{\mu \alpha \mu} (p, 0, -p) p_\alpha  / \left[D(p^2)\right]^2 \, D(0)$. The evaluation of this ratio introduces large statistical
fluctuations which severely limit the estimation of $\Gamma (p^2)$ at high momenta.
Indeed, assuming gaussian error propagation for the estimation of the statistical error on $\Gamma (p^2)$, it follows that
\begin{multline}
   \left[\Delta \Gamma(p^2)\right]^2 = \frac{1}{\left[D(p^2)\right]^4} \Bigg\{
    \left[ \frac{ \Delta G_{\mu \alpha \mu}  p_\alpha } {D(0)}\right]^2 \\ + 
   \left[ 2 ~ \Delta D(p^2) ~  \,  \frac{  G_{\mu \alpha \mu}  p_\alpha } {D(p^2) \, D(0)}\right]^2  \\ + 
   \left[ 2 ~ \Delta D(0) ~  \,  \frac{  G_{\mu \alpha \mu}  p_\alpha } { \left[D(0)\right]^2}\right]^2 \Bigg\} \ ,
\end{multline}   
where $\Delta X$ stands for the statistical error on the quantity $X$.
For large momentum $D(p^2) \sim 1 / p^2$, therefore, and $\Delta \Gamma(p^2) \sim p^4$ and it follows that
the statistical error on $\Gamma (p^2)$ become quite large at higher momentum.
A direct measurement of $\Gamma (p^2)$  with controllable statistical errors is possible only at 
relatively small momentum. \textit{A priori}, if the statistical errors on the various quantities are 
small enough or close to zero, then one can hope to measure the form factor with a relatively good statistical precision. 
However, even in the case of a very large number of gauge configurations, which produces a gluon propagator with
tiny statistical errors, the statistical errors on the three point correlation function dominate and again $\Delta \Gamma (p^2) \sim p^4$. So, the only
function that the lattice can provide for large momentum with controllable statistical errors is the combination 
$\left[D(p^2)\right]^2 \, D(0) \, \Gamma (p^2)$. Certainly, an improvement on the quality of data for ultraviolet region has to go beyond the increase
on the number of configurations. This is a severe limitation of the lattice approach to the evaluation of the three gluon 1PI and it also applies
to the computation of other 1PI with larger number of external legs. 
For the infrared region, the situation is not so dramatic as $D(p^2)$ is essentially constant and
the signal to noise ratio can be improved by considering larger ensembles of gauge configurations.

In order to evaluate possible finite size effects on the lattice computation of $\Gamma (p^2)$, in Fig.~\ref{fig:1PIXXX} we plot the
form factor as a function of the different type of momenta for the two lattices considered here. For the smaller lattice, 
which uses an ensemble with 2000 gauge configurations,  the data for the smallest momenta of type $(n \, 0 \, 0 \, 0)$ are above the remaining 
sets of momenta. For the larger lattice, which uses and ensemble of 279 gauge configurations,
the estimation of $\Gamma (p^2)$ for momenta of type $(n \, 0 \, 0 \, 0)$ provides no valuable information on the form factors.
The $\Gamma (p^2) \, p^2$ data for the simulation using the $80^4$ lattice have large statistical errors and its compatible with a constant value
and the comparison between the two simulations for this type of momenta does not help in the understanding of the quality of the data. 
On the other hand, with the exception of the momenta of type $(n \, 0 \, 0 \, 0)$, the results coming from the two lattices volumes
for $\Gamma (p^2) \, p^2$ are basically the same for the different types of momenta. 
The difference in the statistical errors associated with the two simulations using the $80^4$ and $64^4$ lattices
are due to the difference in the number of configurations in each ensemble.
Further, given that one expects larger finite size effects for momenta of type $(n \, 0 \, 0 \, 0)$, we will ignore the outcome of the simulations associated
with this type momenta from now on.

\begin{figure}[t] %  figure placement: here, top, bottom, or page
   \centering 
   \vspace*{0.5cm}
   \includegraphics[width=3.3in]{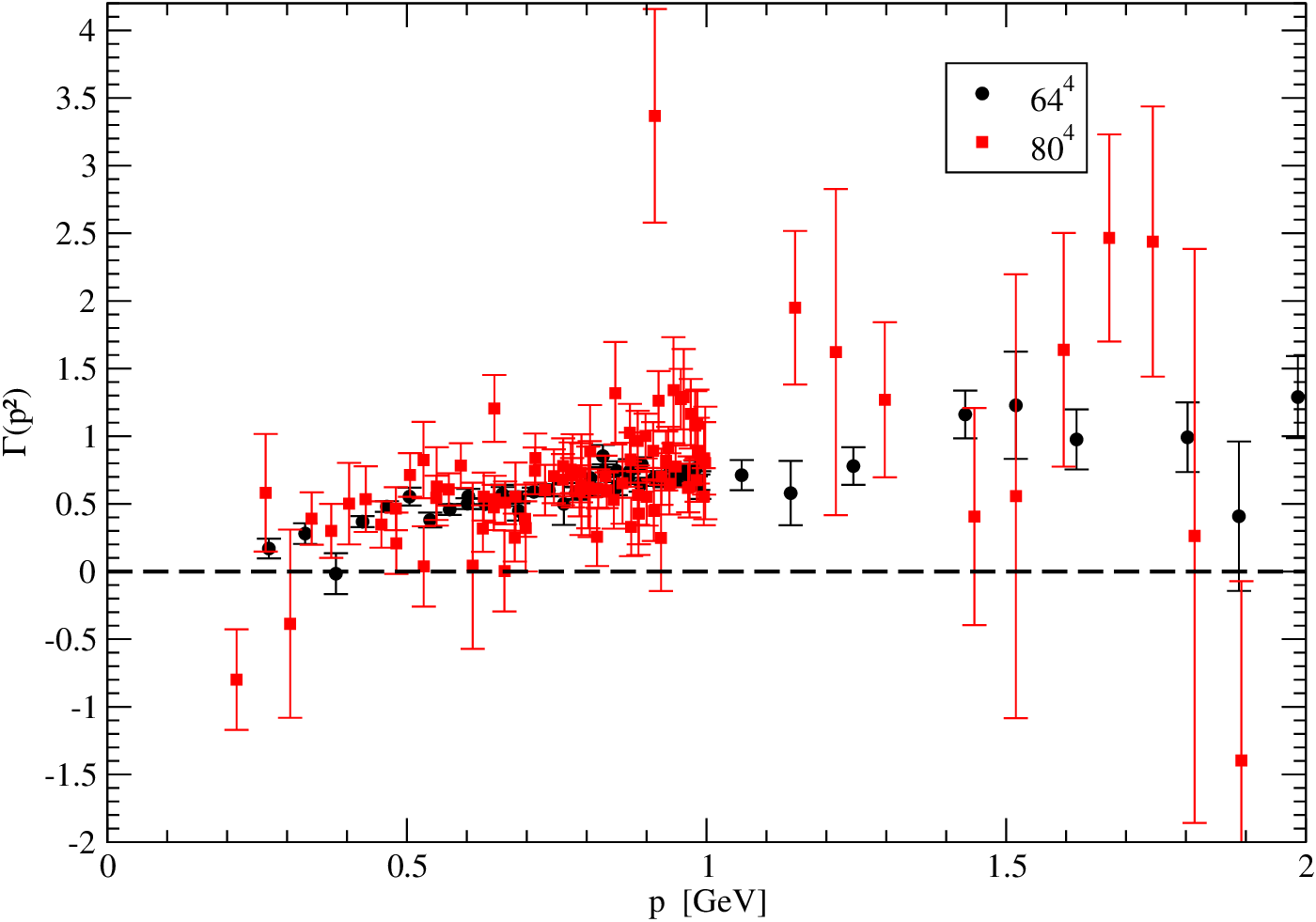} 
   \caption{Low momenta $\Gamma (p^2)$ from the $64^4$ and $80^4$ simulations.}
   \label{fig:1PIgoodall1}
\end{figure}

The form factors $\Gamma (p^2)$ for both data sets and for momenta up to 2 GeV can be seen in Fig.~\ref{fig:1PIgoodall1}. 
As expected, the data for the larger lattice is much noisier than the data for the smaller lattice and this just translates the difference in the size 
of the ensembles used in the calculation.
Despite the larger statistical errors associated to the $80^4$ results, the figures of the two data sets are essentially compatible within one standard deviation. 
As expected and discussed above the statistical errors increase substantially as one approaches the UV region
and, therefore, we have chosen to exclude the "high" momenta in the comparison.
The striking point being that for the lowest available momenta $p = 216$ MeV the form factor describing the three gluon vertex becomes
negative $\Gamma (p^2)$ = -0.80(37) and is compatible with zero only within to 2.2 standard deviations. Further, if one takes into
account the two data points of Fig.~\ref{fig:1PIgoodall1} which correspond to $\Gamma (p = 216 \mbox{ MeV}) = -0.80 (37)$ and
$\Gamma (p = 270 \mbox{ MeV}) = 0.171(73)$ from the simulation with the $64^4$ lattice and
$\Gamma (p = 264 \mbox{ MeV}) = 0.58(43)$ from the simulation with the $80^4$ lattice, the data suggests that the zero crossing for  $\Gamma$ should
occur for momenta  below $\sim 250$ MeV.
The simulations performed for the gauge group SU(2)~\cite{Cucchieri2006,Cucchieri2008}
also suggest that the form factor becomes negative at around the same momentum scale.
The change of sign seen in the recent lattice simulation for SU(3)~\cite{Athenodorou2016} also occurs for momenta within the same range
as observed here. 

\begin{figure}[t] %  figure placement: here, top, bottom, or page
   \centering
   \includegraphics[width=3.3in]{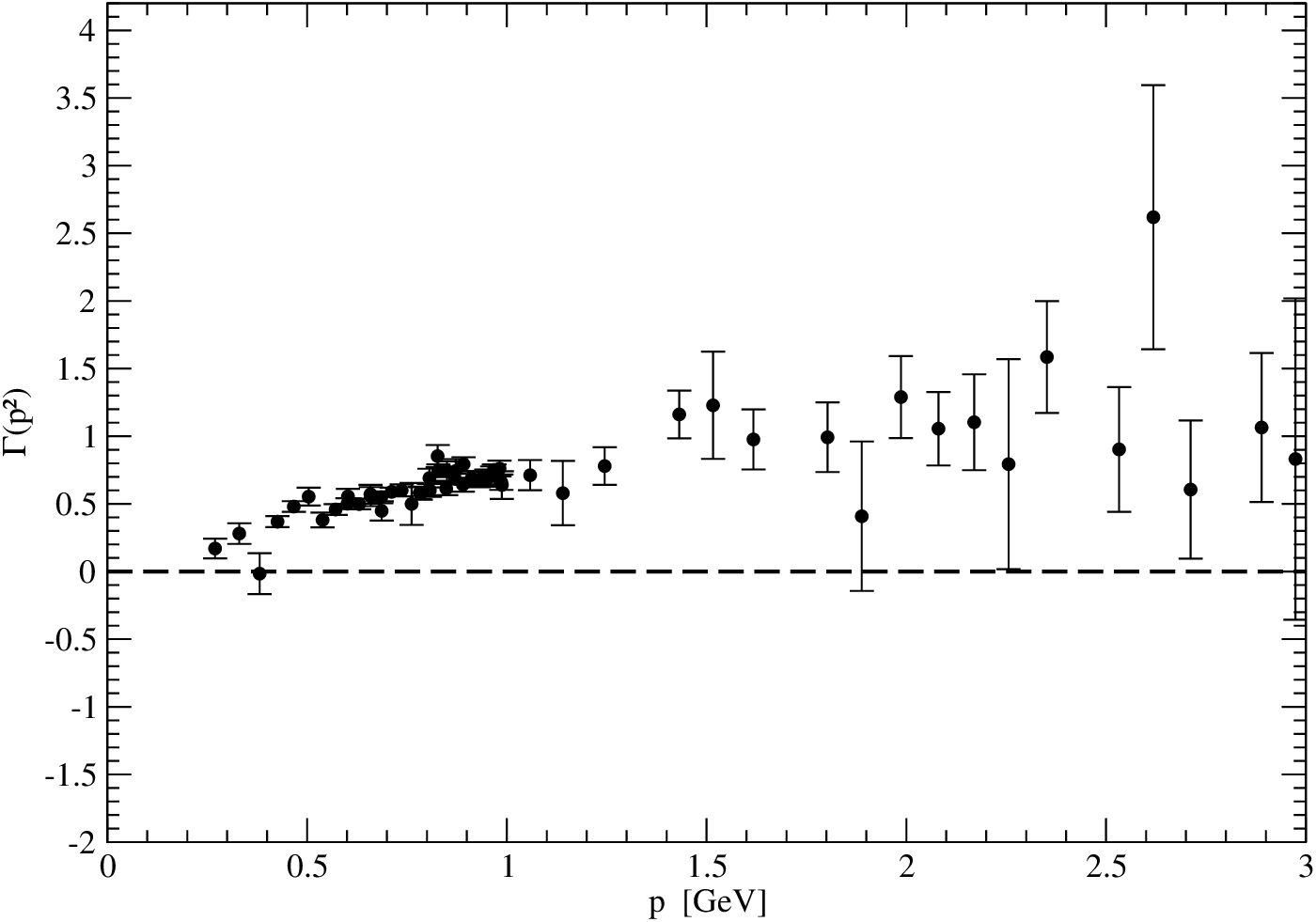} 
   \caption{$\Gamma (p^2)$ from the $64^4$ simulation.}
   \label{fig:1PIgoodall2}
\end{figure}

The computation using the smaller physical volume has small statistical errors, although the zero crossing is not observed. 
In Fig.~\ref{fig:1PIgoodall2} and for completeness we report only the data from the simulation using the $64^4$ data but for an enlarged momentum range.

The study of the ultraviolet behaviour can be performed if one looks to the combination
\begin{equation}
    \Gamma_{UV} (p^2) = \left[ D(p^2) \right]^2 \, D(0) ~ \Gamma (p^2) ~ p^2 \ .
\end{equation}
The one-loop renormalization group improved result for the gluon propagator at high momentum reads
\begin{equation}
    D (p^2) = Z \frac{\left[ \ln \frac{p^2}{\mu^2} \right]^{-\gamma}}{p^2} \ ,
\end{equation}
where $Z$ is a constant, $\mu$ is a renormalization scale and $\gamma = 13/22$ is the gluon anomalous dimension. The equivalent
result for the three gluon 1PI is given by
\begin{equation}
    \Gamma (p^2) = Z^\prime \left[ \ln \frac{p^2}{\mu^2} \right]^{\gamma_{3g}}
\end{equation}
where the anomalous dimension reads $\gamma_{3g} = 17/44$. It follows that for sufficiently high momentum
\begin{equation}
    \Gamma_{UV} (p^2) =  \frac{Z}{p^2} \left[ \ln \frac{p^2}{\mu^2} \right]^{ \gamma_{3g} - 2 \gamma }
    \label{Eq:Pert3glue}
\end{equation}
and $\gamma^\prime =  \gamma_{3g} - 2 \gamma = - 35/44$. 

\begin{figure}[t] %  figure placement: here, top, bottom, or page
   \centering
   \includegraphics[width=3.3in]{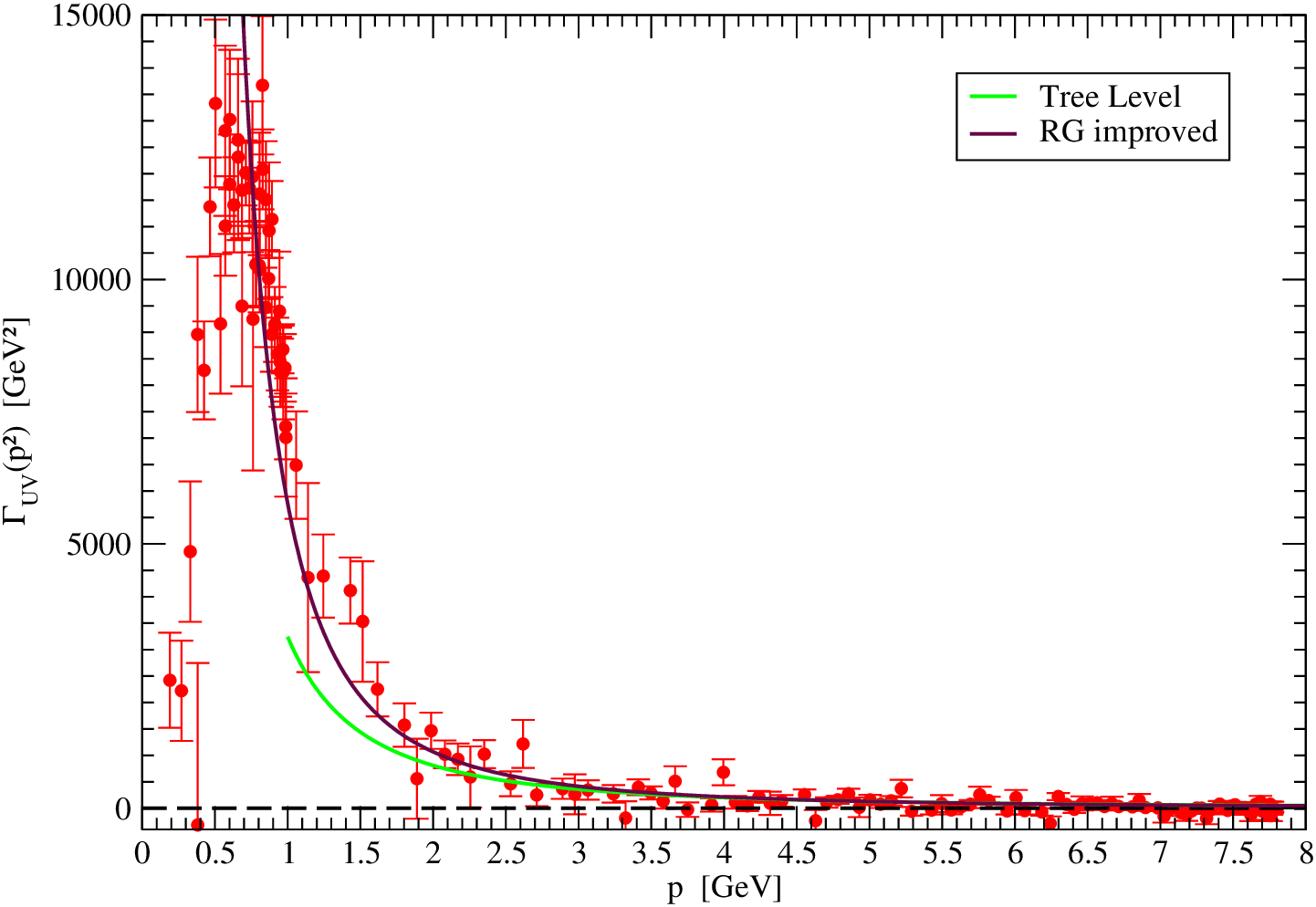} 
   \caption{$\Gamma (p^2)$ from the $64^4$ simulation. The curves are the predictions from perturbation theory. The data points show are those
   which survive the momentum cuts introduced in~\cite{Leinweber1998}. See text for further details.}
   \label{fig:1PIUV}
\end{figure}

The results for $\Gamma_{UV} (p^2)$ measured from the simulation using the $64^4$ lattice and for the full range of momenta 
can be seen in Fig.~\ref{fig:1PIUV}. It includes also the tree level estimation of $\Gamma_{UV} (p^2) = Z/p^2$ and the prediction of Eq. (\ref{Eq:Pert3glue}). 
The normalization constants were set to reproduce the lattice result for momenta $p \sim 5$ GeV and we used for $\mu = 0.22$ GeV to generate the renormalisation improved results. Despite the large statistical errors, the data follows the predictions of perturbation theory for $p$ above $\sim 2.5$ GeV.
This result can be seen as a validation of the perturbative approach to QCD for sufficiently high momenta.

%==================================================================================
%==================================================================================
\section{Summary and Conclusions \label{Sec:Final}}

In the current work we have computed the three gluon complete Green's function on the lattice 
for a particular kinematical configuration and for two different lattices with volume of $(6.5$ fm$)^4$ and
$(8.2$ fm$)^4$. We considered a single value for the lattice spacing $a = 0.102$ fm
and, therefore, one can not estimate possible effects due to the use of a finite $a$.

For the gluon propagator, the results of the two simulations are compatible within one standard deviation and
no finite volume effects are observed. This result is in good agreement with our recent studies
for the gluon propagator~\cite{Oliveira2012,Duarte2016}. On the other hand, the computation
of the 1PI three gluon vertex shows that, in the infrared region, the estimation of $\Gamma (p^2)$ depends on the type 
of momenta considered. This is a clear sign of rotational symmetry breaking effects. The data also suggests that, in the infrared region,
momenta of type $(n \, 0 \, 0 \, 0 )$ over-estimates the 1PI in comparison with the other types of momenta.

The simulations produce a form factor $\Gamma (p^2)$, see Figs.~\ref{fig:1PIgoodall1}, ~\ref{fig:1PIgoodall2} and~\ref{fig:1PIUV}, that is
essentially a constant function at high momenta. For momenta below $\sim 1$ GeV, is a decreasing function of $p^2$ which vanish
at $p \sim 250$ MeV and becomes negative for smaller momenta. The change of sign associated to the three gluon vertex
has also been reported recently in lattice simulations of four dimensional pure Yang-Mills for the gauge group SU(3)~\cite{Athenodorou2016}
and has been reported for three dimensional pure Yang-Mills for the gauge group SU(2) in~\cite{Cucchieri2006,Cucchieri2008}.
All the lattice simulations point towards the same momentum scale for the zero crossing of the three gluon vertex. 
As discussed in~\cite{Binosi2013,Aguilar2014} the observed change of sign in the three gluon vertex seems to be a requirement to have a
properly defined Dyson-Schwinger equation for the gluon propagator.

In what concerns the high momentum region, the lattice data is compatible with the prediction of the
renormalisation group improved perturbation theory.
In this sense, our data validates the perturbative approach to QCD for the UV region.

In the current work, we also discussed the limitation of the lattice approach to the evaluation of 1PI for the case of multiple gluon legs.
Our conclusion being that, within the approach considered here, the UV region will be always plagued with large statistical errors and, therefore,
the method is not able to access the high momentum region in a way to extract reliable information. Instead, for higher momenta one can
consider combinations of the 1PI and correlation functions with smaller number of external legs. Combined with a very accurate estimation
of these later functions one can, in principle, still get some usual information about the UV behaviour of the 1PI. An accurate estimation of the
1PI in the UV region clearly requires going beyond the increase of the size of configuration ensembles.

 %==================================================================================
  %==================================================================================
 \begin{appendix}
 
 %==================================================================================
  %==================================================================================
 \section{Kinematical Decomposition of $\Gamma_{\alpha\beta\gamma}$ \label{Sec:GammaBallChiu}}
 
For completeness, here we describe the tensor basis introduced in~\cite{BallChiu1980} 
to describe the  $\Gamma_{\mu_1 \mu_2 \mu_3} (p_1, p_2, p_3)$ function of the 1PI three gluon vertex defined in Eq. (\ref{Eq:gammaLorentz}).  
In Minkowsky space, the transverse component of the vertex parts reads
\begin{widetext}
\begin{multline} 
 \Gamma^{(t)} _{\mu_1 \mu_2 \mu_3} (p_1, p_2, p_3) = 
 F(p^2_1, p^2_2; p^2_3) \Big[ g_{\mu_1\mu_2} \, (p_1 \cdot p_2) - {p_1}_{\mu_2} ~ {p_2} _{\mu_1} \Big] \, B^3_{\mu_3} \\
 + H(p^2_1, p^2_2, p^2_3) \left[ -g_{\mu_1\mu_2}\, B^3_{\mu_3} + \frac{1}{3} \big( {p_1}_{\mu_3} ~ {p_2}_{\mu_1} ~ {p_3}_{\mu_2} 
                                                                                           - {p_1}_{\mu_2} ~ {p_2}_{\mu_3} ~ {p_3}_{\mu_1}  \big) \right] \\
                                                                                            + ~\mbox{ cyclic permutations}
 \label{3g_trans}
\end{multline} 
\end{widetext}
where
\begin{equation}
   B^3_{\mu_3} = {p_1}_{\mu_3} \, \left( p_2 \cdot p_3  \right) - {p_2}_{\mu_3} \, \left( p_1 \cdot p_3 \right)\, ,
\end{equation}
$F(p^2_1, p^2_2; p^2_3)$ is a scalar function symmetric under interchange of the first two arguments and
$H(p^2_1, p^2_2, p^2_3) $ is totally symmetric under interchange of its arguments. It follows from the definition that
\begin{equation}
  p_3^{\mu_3} \, \Gamma^{(t)} _{\mu_1 \mu_2 \mu_3} (p_1, p_2, p_3) = 0
\end{equation}  
and Bose symmetry of the vertex ensures that the contractions with $p_1$ and $p_2$ also vanishes.
On the other hand, the longitudinal part of the vertex is given by
\begin{widetext}
\begin{multline} 
 \Gamma^{(l)} _{\mu_1 \mu_2 \mu_3} (p_1, p_2, p_3) = 
 A(p^2_1, p^2_2; p^2_3) ~ g_{\mu_1\mu_2} \, \big(  {p_1} -  {p_2}  \big)_{\mu_3} ~
 + ~ B(p^2_1, p^2_2; p^2_3) ~ g_{\mu_1\mu_2} \, \big(  {p_1} +  {p_2} \big)_{\mu_3} \\
 + C(p^2_1, p^2_2; p^2_3) \, \big(  {p_1}_{\mu_2}  {p_2} _{\mu_1} - g_{\mu_1 \mu_2} \, p_1 \cdot p_2 \big)  \, \big(  {p_1} -  {p_2}  \big) _{\mu_3} \\
 + \frac{1}{3} S(p^2_1, p^2_2, p^2_3) \big( {p_1}_{\mu_3} ~ {p_2}_{\mu_1} ~ {p_3}_{\mu_2}  +    {p_1}_{\mu_2} ~ {p_2}_{\mu_3} ~ {p_3}_{\mu_1} \big)
 \\ + ~\mbox{ cyclic permutations}
 \label{3g_long}
\end{multline} 
\end{widetext}
where the scalar functions $A(p^2_1, p^2_2; p^2_3)$ and $C(p^2_1, p^2_2; p^2_3)$ are symmetric in their first two arguments,
$B(p^2_1, p^2_2; p^2_3)$ is antisymmetric and $S(p^2_1, p^2_2, p^2_3)$ is antisymmetric under exchange of any pair of arguments.
The full 1PI reads
\begin{multline}
 \Gamma _{\mu_1 \mu_2 \mu_3} (p_1, p_2, p_3) = \Gamma^{(t)} _{\mu_1 \mu_2 \mu_3} (p_1, p_2, p_3)  \\ +
  \Gamma^{(l)} _{\mu_1 \mu_2 \mu_3} (p_1, p_2, p_3) \ .
\end{multline}

On the lattice one computes the Green function (\ref{Eq:GLat}) on a reduced set of momentum configurations and from
lattice simulations can only access combinations of the form factors  $F$, $H$, $A$, $B$, $C$, $S$
on a finite set of momentum configurations. 

\end{appendix}

\begin{acknowledgments}
The authors acknowledge the Laboratory for Advanced Computing at University of Coimbra for providing HPC computing resources 
Navigator that have contributed to the research results reported within this paper (URL http://www.lca.uc.pt). 
This work was granted access to the HPC resources of the PDC Center for High Performance Computing at the KTH Royal Institute of Technology, Sweden, 
made available within the Distributed European Computing Initiative by the PRACE-2IP, receiving funding from the 
European Community's Seventh Framework Programme (FP7/2007-2013) under grand agreement no. RI-283493. The use of Lindgren has been provided 
under DECI-9 project COIMBRALATT.
We acknowledge that the results of this research have been achieved using the PRACE-3IP project (FP7 RI-312763) resource Sisu based in Finland at CSC. 
The use of Sisu has been provided under DECI-12 project COIMBRALATT2.
P. J. Silva acknowledges partial support by FCT under Contract No. SFRH/BPD/40998/2007.
\end{acknowledgments}

\end{document}